\documentclass[aps,prb,twocolumn,
superscriptaddress,onlinecite]{revtex4-1}

\usepackage{graphicx}
\usepackage{epstopdf}
\usepackage[scientific-notation=false]{siunitx}

\begin{document}


\title{Surface state reconstruction in ion-damaged SmB$_{6}$}


\author{N. Wakeham}
\affiliation{Los Alamos National Laboratory}

\author{Y. Q. Wang}
\affiliation{Los Alamos National Laboratory}

\author{Z. Fisk}%
\affiliation{Dept. of Physics and Astronomy, University of California, Irvine}%

\author{F. Ronning}
\affiliation{Los Alamos National Laboratory}

\author{J. D. Thompson}
\affiliation{Los Alamos National Laboratory}


\date{\today}

\begin{abstract}
We have used ion-irradiation to damage the (001) surfaces of SmB$_6$ single crystals to varying depths, and have measured the resistivity as a function of temperature for each depth of damage. We observe a reduction in the residual resistivity with increasing depth of damage. Our data are consistent with a model in which the surface state is not destroyed by the ion-irradiation, but instead the damaged layer is poorly conducting and the initial surface state is reconstructed below the damage. This behavior is consistent with a surface state that is topologically protected.
\end{abstract}


\maketitle
\section{Introduction}
Topological insulators (TI) have a set of bulk wavefunctions that are topologically distinct from the vacuum. One consequence of this is at the interface between the topologically non-trivial insulator and the trivial vacuum the energy gap must close \cite{Ando2013}. This gives rise to a gapless surface state that in a time reversal invariant system is said to be topologically protected because it is robust against time-reversal invariant perturbations. TI have been intensively studied because of their ability to further our understanding of the importance of the topology of wave-functions, but also because properties of the surface state, for example spin-momentum locking, have practical applications in areas such as spintronics \cite{Yokoyama2014}.

SmB$_6$ has long been established as a Kondo insulator, in which the hybridization of the $f$ electrons with the conduction electrons opens a small energy gap \cite{Kasuya1994}. This energy gap causes the resistivity to diverge with decreasing temperature down to 5\,K. However, below 5\,K the resistivity saturates. Recently, SmB$_6$ was predicted to be topologically non-trivial, and therefore a three-dimensional topological Kondo insulator with a topologically protected surface state \cite{Dzero2010}. Experiments by several groups have established strong evidence that the resistivity saturation is indeed due to a conductive and intrinsic surface state \cite{Syers2014,Wolgast2013,Kim2013}. Furthermore, transport measurements have shown weak anti-localization \cite{Thomas2013} and the sensitivity of the surface state to time reversal breaking perturbations \cite{Kim2014} that are consistent with the surface state being topological. Angle resolved photoemission spectroscopy (ARPES) experiments have also shown evidence for a topological surface state through observation of in-gap surface states and spin-momentum locking \cite{Neupane2013,Xu2014}. However, the topological nature of the surface state has not been conclusively proven. There is evidence from scanning tunneling spectroscopy measurements for surface reconstructions \cite{Robler2014}, and x-ray measurements have indicated the physical surface of SmB$_6$ crystals has a different average valence of Sm ions compared with the bulk \cite{Aono1979,Phelan2014}. These factors may give rise to a non-topological conducting surface state, perhaps as well as a topological one. Hence, the properties of this surface state are still an area of interest.

Here, we report on the nature of the intrinsic surface state in SmB$_6$ through non-magnetic ion-irradiation of the surface of SmB$_6$ single crystals. We have measured the temperature dependent resistivity of two samples as a function of the depth to which the surface is damaged, and find that the intrinsic surface state is not destroyed by this time-reversal invariant damage. Instead, the damaged area becomes more metallic than bulk SmB$_6$, and the intrinsic surface state is reconstructed below this damaged layer. Such insensitivity to this time-reversal invariant perturbation is consistent with the surface state in SmB$_6$ being topologically protected.

\section{Experimental Techniques}
Single crystals of SmB$_6$ were grown using aluminum flux and then polished to give plate-like crystals in the (001) direction with approximate dimensions of the length, width and thickness of \SI{300}{\micro\metre}, \SI{180}{\micro\metre} and \SI{35}{\micro\metre}, respectively. The resistance $R$ of the samples as a function of temperature $T$ was measured using a four-probe low frequency AC resistance bridge with spot-welded contacts of \SI{25}{\micro\metre} platinum wires attached to the top (001) face. These contacts remained in place during all of the measurements and ion-irradiation. Temperature was controlled within a Quantum Design PPMS. Note that the plate-like geometry of the samples means that the sides of the samples are not a significant conducting channel compared with the larger faces, and are therefore ignored in our discussions. The (001) surfaces of the samples were damaged by ion-irradiation. The ion used, acceleration energy, ion fluence and exposure time to produce a peak damage of at least 1 displacement per atom (DPA) are shown in Table \ref{Ion details}. The level of damage as a function of depth for each exposure was calculated using the as SRIM Monte Carlo code in the full cascade mode, and is shown in the inset to Fig. \ref{RvsT}  \cite{Ziegler2010}. The damage depth is defined as the depth at which there is half of the maximum damage.
\begin{table}[ht]

\centering 
{
\renewcommand{\arraystretch}{1.4}
\begin{tabular}{c c c c c} 
\hline\hline 
Depth [nm]& Ion & Energy [keV] & Ion fluence [cm$^{-2}$]& Time [s]\\ [0.5ex] 
\hline 
12.9 & Xe$^+$ & 20 & \SI{5.0E14}  & 800 \\ 
108 & Xe$^{2+}$ & 350 & \SI{3.0E14} & 1034 \\
237 & Ar$^{2+}$ & 300 & \SI{1.5E15} & 357 \\ [1ex] 
\hline 
\end{tabular}
}

\caption{Parameters used in the ion-irradiation of SmB$_6$ to produce a damage profile of approximately 1 displacement per atom down to the stated depth.} 
\label{Ion details} 
\end{table}

\section{Results}
\begin{figure}[h]
 \includegraphics[width=0.99\columnwidth]{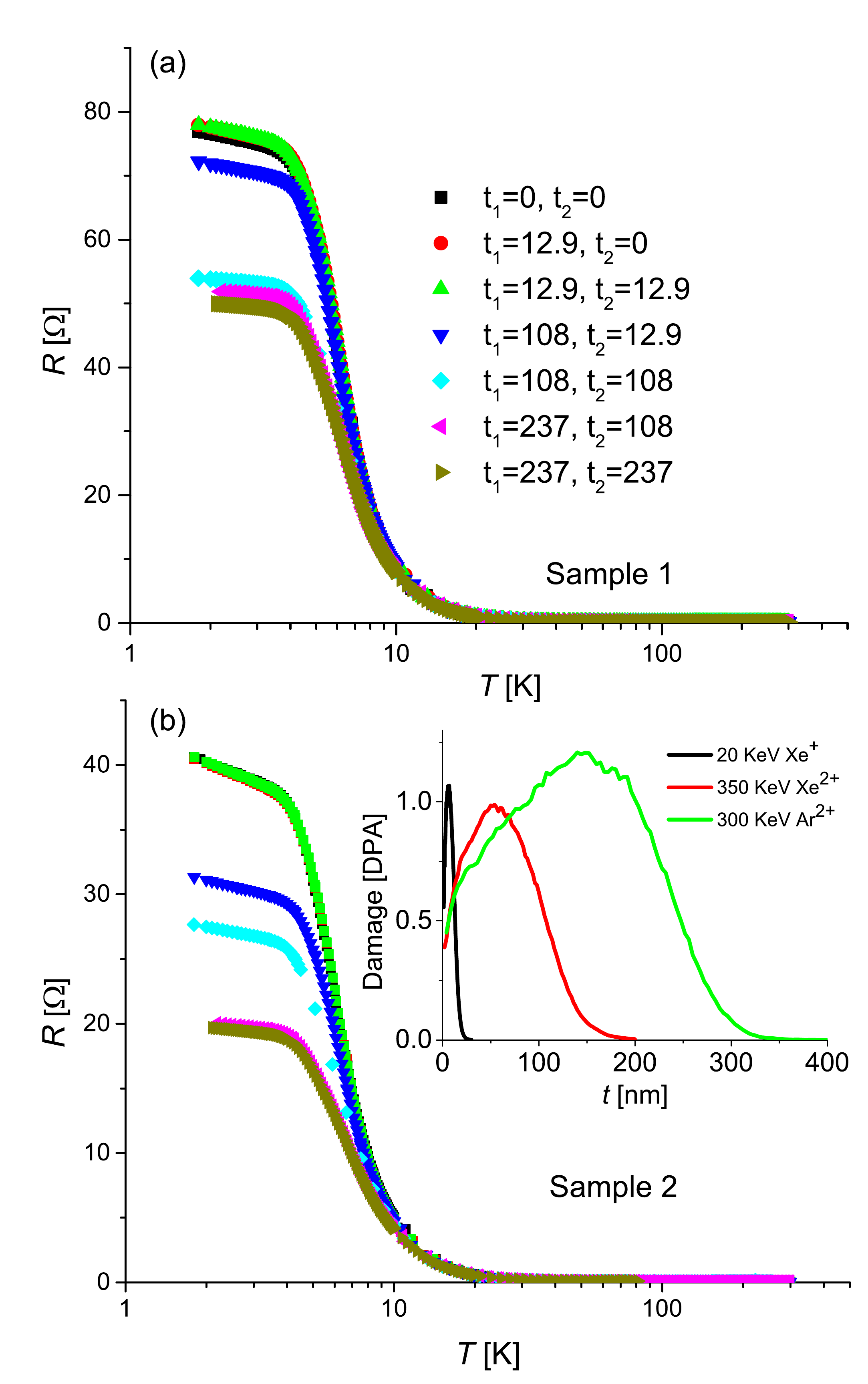}%
 \caption{(Color online) Temperature dependence of the resistance of SmB$_6$ after differing depths of ion-irradiation damage to each face. $t_1$ is the depth of damage to the top face of the sample (to which contacts were made), $t_2$ is the depth of damage to the bottom face, both in units of nm. Shown for (a) sample 1 and (b) sample 2. The inset shows the calculated damage as a function of depth for each ion beam exposure described in Table \ref{Ion details}.\label{RvsT}}
 \end{figure}

The temperature dependent resistance of the samples was measured after being damaged to the stated depth on the top face, and then again after being damaged on the bottom face. $R(T)$ as the depth of damage was varied on each face is shown in Fig. \ref{RvsT}.
This figure clearly shows a marked reduction in the low temperature residual resistance $R_0$ as the damage depth was increased. A fit to the data (not shown) based on an activation-like resistance at high temperature and temperature independent surface resistance at low temperature, discussed by Kim \emph{et al.} \cite{Kim2013}, indicates no significant change to the bulk energy gap of the system after the surfaces were damaged. Therefore, the reduction in $R_0$ suggests that either the damage increased the conductivity of the intrinsic surface state, or it introduced a new conducting channel. Karkin \emph{et al.} irradiated SmB$_6$ with neutrons to produce damage of 10 DPA throughout the sample. This damage reduced the residual resistivity by 3 orders of magnitude, and gave a largely temperature independent resistivity \cite{Karkin2007}. Even modest damage to the bulk of SmB$_6$ by neutrons reduced the low temperature resistivity \cite{Irradids1983}. The shapes of the defects induced by low energy ion-irradiation and neutrons are expected to be qualitatively similar \cite{Was2014}. We conclude that in our measurements one effect of the ion irradiation was to create a thin damaged  layer of relatively low resistivity that reduced the overall resistance of the sample at low temperatures. In addition to this, we must also determine whether the intrinsic surface state present in the undamaged sample was destroyed by the damage or was reconstructed below it.

\section{Discussion}
\begin{figure}[h]
 \includegraphics[width=0.9\columnwidth]{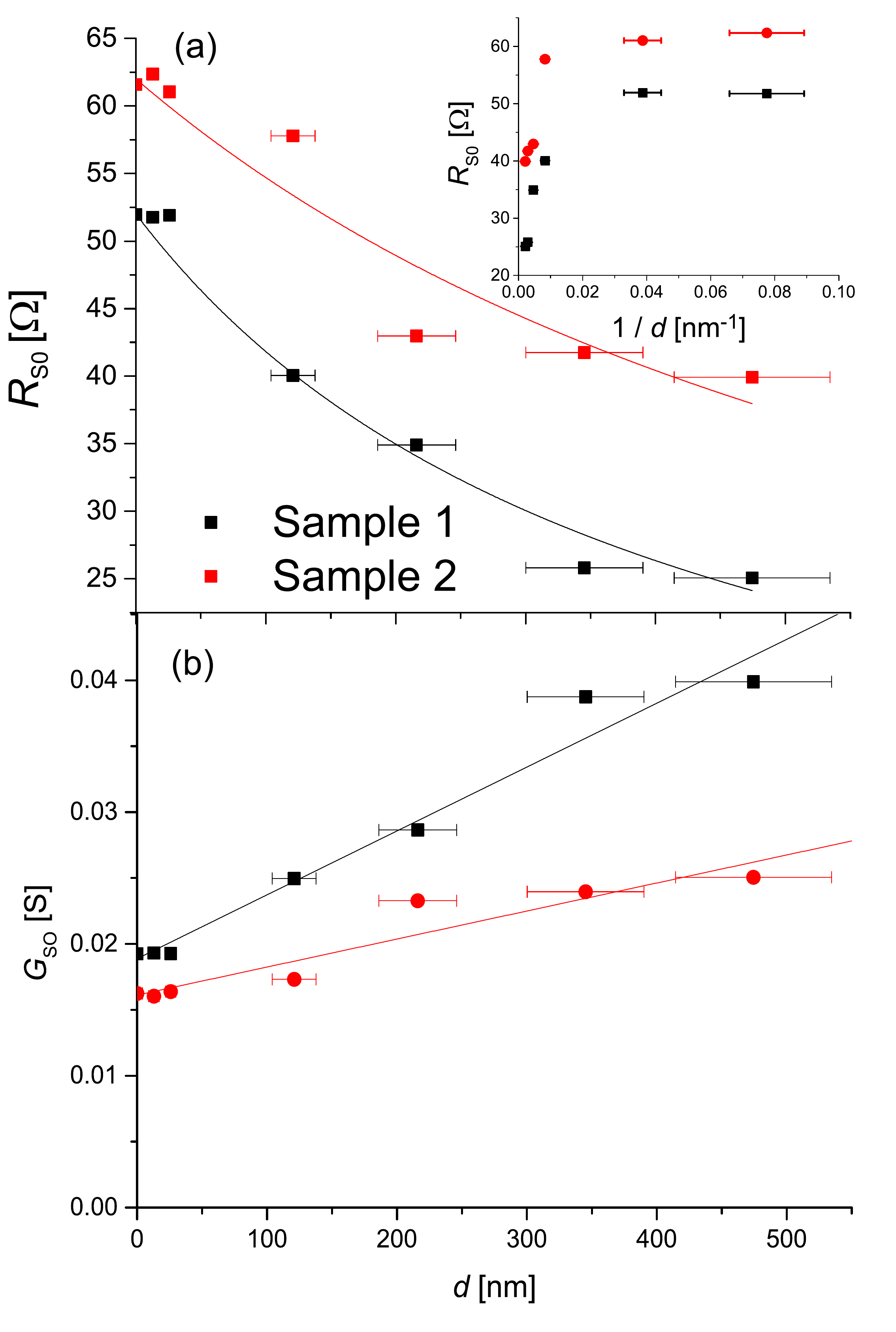}%
 \caption{ (Color online) (a) Residual sheet resistance $R_{S0}$ and (b) residual sheet conductance $G_{S0}$ of two samples of SmB$_6$ as a function of the effective depth of ion-radiation damage $d$, given by the sum of the damage depth to the top and bottom faces. Solid lines are fits to the formula $1/R_{S0} = G_{S0}= 1/R_{SS} +d/2\rho_{DL}$, where $\rho_{DL}$ is the resistivity of the damaged layer. Inset shows $R_{S0}$ as function of $1/d$ to emphasize the lack of proportionality between the two. \label{Rvs1overd}}
 \end{figure}

To interpret the change in low temperature resistance, it is useful to convert the residual resistance to a sheet resistance $R_{S0}$, given by $R_{S0}=R_0 2w/l$, where $w$ is the width of the sample and $l$ is the distance between voltage contacts. Fig. \ref{Rvs1overd}a shows $R_{S0}$ as a function of the total depth of surface damage $d$, where $d$ is the sum of the depth of damage of the top and bottom faces.
The inset of Fig. \ref{Rvs1overd}a shows $R_{S0}$ as a function of $1/d$ and demonstrates there is not a $1/d$ dependence. The residual sheet conductance of the sample is given by $G_{S0} = 1/R_{S0}$, and is plotted against $d$ in Fig. \ref{Rvs1overd}b. The solid curves in Fig. \ref{Rvs1overd}a and \ref{Rvs1overd}b show that the $d$ dependence can be described by a model of reconstruction of the surface state represented by two parallel resistors.

\begin{figure}[h]
 \includegraphics[width=0.9\columnwidth]{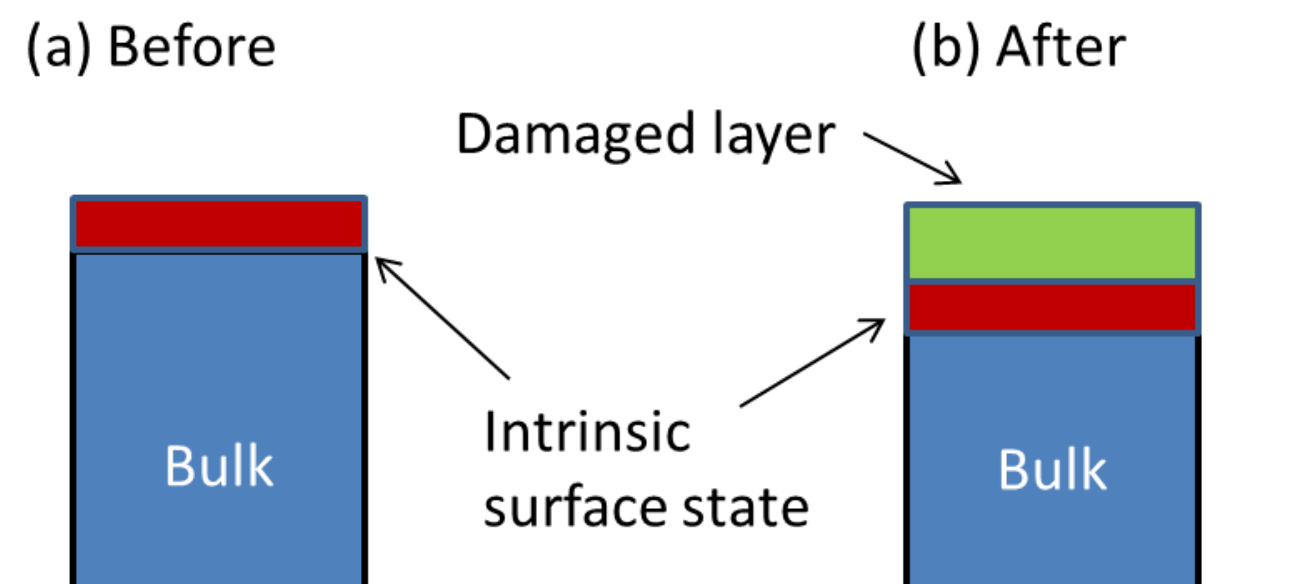}%
 \caption{ (Color online) Schematic diagram of the top surface of SmB$_6$ (a) before and (b) after ion-irradiation damage. This shows the model of reconstruction of the intrinsic surface state (red) underneath a conductive damaged layer (green). \label{SurfaceRecon}}
 \end{figure}

A schematic diagram of the parallel resistor scenario is shown in Fig. \ref{SurfaceRecon}.
The first resistor represents the intrinsic surface state, and is assumed to have a constant sheet resistance $R_{SS}$, as function of $d$. This constant is given by $R_{S0}$ in the undamaged sample. The second resistor represents the damaged layer, and therefore the sheet resistance is given by $2\rho_{DL}/d$, where $\rho_{DL}$ is the resistivity of the damaged layer, which is a fitted parameter. The bulk is assumed to be perfectly insulating. The total sheet resistance $R_{S0}$ is therefore given by
\begin{equation}
\frac{1}{R_{S0}}= G_{S0} = \frac{1}{R_{SS}} +\frac{d}{2\rho_{DL}}
\end{equation}
 The fitted parameter $\rho_{DL}$  is $1.0\pm 0.2$\,\SI{}{\milli\ohm\centi\metre} and $2.4\pm 0.4$\,\SI{}{\milli\ohm\centi\metre} in sample 1 and 2, respectively. This is in reasonable agreement with the resistivity of \SI{1.5}{\milli\ohm\centi\metre} found in the heavily neutron damaged sample \cite{Karkin2007}, as is expected since in both cases the irradiated region is sufficiently damaged to be considered approximately amorphous.




The origin of the reduction in magnitude of dR/dT with increasing depths of damage at temperatures below 4\,K shown in Fig. \ref{RvsT} cannot be ascertained from our data. It may arise from a temperature dependent resistivity of the damaged layer, or a change in the temperature dependence of the intrinsic surface state. The data shown in Fig. \ref{Rvs1overd} demonstrate deviations from the fitted line that are seemingly beyond the expected statistical error associated with the uncertainty in the depth of damage. This may arise from inhomogeneity of the damage to the sample, particularly as a function of $d$. In addition, there may also be changes to the conductivity of the intrinsic surface state not considered in our model. Note that this would be particularly significant for small $d$.

Let us now consider scenarios for the effect of the damage that are alternatives to the surface state reconstruction discussed above. Firstly, note that the inset of Fig. \ref{Rvs1overd} shows that the sheet resistance does not follow a $1/d$ dependence that would be expected if the intrinsic surface state depth were very shallow ($<$ 10 nm), and this were destroyed by the ion irradiation and replaced with the conducting damaged layer. We therefore exclude this possibility.

The depth of the surface state $\xi$ may be crudely estimated from measurements of the Fermi velocity $v_F$ of the surface state using $\xi=\hbar v_F/\Delta$, where $\Delta$ is the magnitude of the Kondo energy gap, taken to be 38$k_B$\,J, and $k_B$ is the Boltzmann constant \cite{Kim2013}.  Using this equation, the value of $v_F\approx 6$ eV$\rm{\AA}$ obtained from de Haas-van Alphen (dHvA) measurements implies a metallic surface depth of  $\sim 150$ nm \cite{Li2014a}. Therefore, we must consider another scenario, that the intrinsic surface state has a depth of around $100$ nm, and is destroyed by the ion damage. In this picture, ion irradiation damage to 12.9 nm may have little effect on the sample resistance. However, once damaged to 108 nm the surface state could be destroyed, and then at larger $d$ the resistance as a function of $1/d$ would follow a linear dependence. The calculated resistivity of the damaged layer obtained by fitting R$_{S0}$ for  $d>108$ nm with a $1/d$ dependence is of order \SI{100}{\micro\ohm\centi\metre}. This resistivity value is more than an order of magnitude smaller than that measured for the neutron damaged SmB$_{6}$ \cite{Karkin2007}. In addition, the room temperature bulk resistivity of the measured crystals was $\sim$\SI{150}{\micro\ohm\centi\metre}. It therefore seems unlikely that the heavily damaged material could support such a low resistivity. Work by others also argues against the surface state having a thickness of $\sim 100$\,nm. For instance, the values of $v_F\approx300$\,meV$\rm{\AA}$ obtained from ARPES measurements are much smaller than obtained through dHvA measurements \cite{Neupane2013, Jiang2013}. This smaller value implies the penetration of the surface state is $\xi\approx10$\,nm. We also note that measurements of 100 nm thick films of SmB$_{6}$ show a temperature dependence like that of bulk crystals, and is argued to reflect the development of a Kondo-derived gap, and a saturated resistivity associated with a conductive surface state \cite{Yong2014}. Because some fraction of the film's volume hosts the Kondo-gap, the penetration depth of the surface state must be well below 50\,nm.  Theoretical estimates of the depth of the topological surface state are $1-25$ lattice spacings or $0.4-10$\,nm \cite{Roy2014, Alexandrov2013, Alexandrov2015}. Based on all of these factors we conclude that a surface state of 100 nm and a highly conducting damaged layer are unlikely to be the explanation of our results.

A final scenario to consider in the explanation of the residual resistance as a function of damage depth is that the undamaged sample had a topologically trivial insulating surface layer greater than 200 nm deep, with the intrinsic surface state below that. The layer would most likely be the oxide that forms on SmB$_6$ crystals in air, and is known to be an insulator \cite{Tanaka1980}. The presence of such a deep layer would mean that the irradiation damage never went deep enough to affect the intrinsic surface state, it merely converted an insulating surface layer into a poor metallic layer, and therefore produced a reduced residual resistance. However, such an oxide layer would be expected to form to a depth of order 10 unit cells or approximately 4\,nm, and indeed has recently been reported to be 2\,nm deep in SmB$_{6}$ \cite{Li2014}. This is much less than the 200 nm required for this scenario. It is also unlikely that the highly insulating oxide layer would be made conducting by ion-irradiation. Hence, we rule out the oxide layer as the explanation of our results. 

\section{Conclusion}
Given all the considerations discussed above, we conclude that ion irradiation of the surface of SmB$_6$ produced a damaged layer with a resistivity of around \SI{1.5}{\milli\ohm\centi\metre}, but that the intrinsic conductive surface state present in the undamaged sample was reconstructed below the damaged layer. To our knowledge there are no theoretical predictions for the interface between a topological insulator and a conducting amorphous material. However, it was shown theoretically that the interface between a crystalline metal and TI will retain the Dirac spectrum expected at the interface to a vacuum, if the orbital overlap between the metal and the TI is small relative to the hopping parameter of the metal \cite{Zhao2010}. In our case, the large disorder in the metallic damaged layer means the resistivity is large and close to the Anderson localization limit, and therefore there is likely to be little orbital overlap with the TI. One further consideration is that ion-irradiation causes a distribution of disorder as a function of depth into the sample. Therefore, the surface state is likely to have formed within a partially disordered region of the sample. Roy \emph{et al.} have calculated that this may shift the Dirac cone of the surface state relative to the valence band, but will not destroy the surface state \cite{Roy2014}. Hence, we conclude that our deduction of a reconstructed surface state is consistent with the theoretical predictions for a topological surface state in SmB$_{6}$.

Our results may be relevant to the study of heterostructures of SmB$_6$ combined with insulators, superconductors and normal metals that have been proposed \cite{Hou2012,Affleck2013,Williams2012}. We show that although the surface state will likely be present in such structures, the surface state may reside deeper within the SmB$_6$ layer than anticipated. In the future it will be interesting to compare our results with the effect of implanting magnetic ions into the surface of SmB$_6$. One might expect this to break the time reversal symmetry at the surface and therefore destroy the intrinsic surface state, as was shown by the introduction of Gd ions into the bulk of SmB$_6$ \cite{Kim2014}, and the deposition of iron onto the surface of Bi$_2$Se$_3$ \cite{Wray2010}.


NW acknowledges the support of the Los Alamos National Laboratory LDRD program.  The work of FR and JT was performed under the auspices of the U.S. Department of Energy, Office of Science. The Ion implantation facility was partially supported by the Center for Integrated Nanotechnologies (CINT), a DOE nanoscience user facility jointly operated by Los Alamos and Sandia National Laboratories.
\bibliography{MyCollection, MyLibrary}

\end{document}